\documentclass[11pt]{article}
\usepackage[utf8]{inputenc}
\usepackage{booktabs}
\usepackage{hyperref}
\usepackage{graphicx}
\usepackage{color}
\usepackage{caption}
\usepackage{lineno}
\usepackage{comment}
\usepackage{aas_macros}
\usepackage{amssymb}
\usepackage{verbatim}

\title{Discovery of Strongly-lensed Gravitational Waves -- Implications for the LSST Observing Strategy}

\author{
Graham P. Smith, Andrew Robertson, Matteo Bianconi, and  
Mathilde Jauzac\\ on behalf of the Strong Lensing Science Collaboration and \\ the Gravitationally Lensed Gravitational Wave Hunters}

\date{November 2018}

\begin{document}

\maketitle

\begin{abstract}
LSST's wide-field of view and sensitivity will revolutionize studies
of the transient sky by finding extraordinary numbers of new
transients every night.  The recent discovery of a kilonova
counterpart to LIGO/Virgo's first detection of gravitational waves
(GWs) from a double neutron star (NS-NS) merger also creates an
exciting opportunity for LSST to offer a Target of Opportunity (ToO)
mode of observing.  We have been exploring the possibility of
detecting strongly lensed GWs, that would enable new tests of GR,
extend multi-messenger astronomy out to $z\gtrsim1$, and deliver a new
class of sub-millisecond precision time-delay constraints on lens mass
distributions.  We forecast that the rate of detection of lensed NS-NS
mergers in the 2020s will be $\sim0.1$ per Earth year, that the
typical source will be at $z\simeq2$, and that the multiply-imaged
kilonova counterpart will have a magnitude of ${\rm AB}\simeq25.4$ in
$g/r/i$-band filters -- i.e. fainter than the sensitivity of a single
LSST WFD visit.  We therefore advocate (1) creating a flexible and
efficient Target of Opportunity programme within the LSST observing
strategy that is capable of discovering sources fainter than
single-visit depth, and (2) surveying the entire observable
extragalactic sky as rapidly as possible in the WFD survey.  The
latter will enable a very broad range of early science that relies on
wide survey area for detection of large samples of objects and/or
maximizing the fraction of sky over which reference imaging is
available.  For example, it will enable prompt discovery of a uniform
and all-sky sample of galaxy/group/cluster-scale lenses that will
underpin LSST strong-lensing science.  This white paper complements
submissions from DESC, SLSC, and TVSSC, that discuss kilonova, GW, and
strong lensing.

\end{abstract}

\section{White Paper Information}

\noindent  
Corresponding authors:\\
Graham Smith, gps@star.sr.bham.ac.uk\\ 
Matteo Bianconi, mbianconi@star.sr.bham.ac.uk\\
Mathilde Jauzac, mathilde.jauzac@durham.ac.uk\\
Andrew Robertson, andrew.robertson@durham.ac.uk\\

\noindent
This White Paper has been developed in consultation with:\\
Aprajita Verma, aprajita.verma@physics.ox.ac.uk, SLSC Co-chair\\
Tom Collett, thomas.collett@port.ac.uk, DESC-SLWG Co-convenor\\

\noindent
This White Paper has benefited from discussions with:\\
Christopher Berry (Northwestern), Will Farr (Stony Brook), Richard Massey (Durham), Johan Richard (Lyon), Dan Ryczanowski (Birmingham), Keren Sharon (Michigan), John Veitch (Glasgow), Alberto Vecchio (Birmingham).

\begin{enumerate} 
\item {\bf Science Category:}
\begin{itemize}
\item The Nature of Dark Matter and Understanding Dark Energy
\item Exploring the Changing Sky
\end{itemize}

\item {\bf Survey Type Category:} 
Wide-Fast-Deep, Target of Opportunity observations\\

\item {\bf Observing Strategy Category:}
  \begin{itemize} 
  \item other category: This paper seeks to highlights primarily the
    observing conditions constrains for the potential for target of
    opportunity follow-up of lensed gravitational wave sources. It is
    largely agnostic to the specifics of the short-term observing
    strategy to build up exposures for the WFD to full depth in
    general, but does support securing a full survey of the entire sky
    to single visit depth as soon as possible after the start of LSST
    operations.
    \end{itemize}  
\end{enumerate}

\clearpage

\section{Scientific Motivation}

Discovery of the first strongly-lensed (i.e. multiply-imaged)
gravitational wave (GW) will bring together two of the central
predictions of General Relativity (GR) -- gravitational waves and
gravitational lensing.  A single lensed GW source will therefore be a
huge scientific breakthrough and will enable a broad range of exciting
science, including: the first glimpse of the stellar mass compact
object binaries at $z>1$, novel tests of GR aided by an increase in
the number of GW detectors thanks to multiple detections of the same
source, a new class of constraints on the mass distribution in
galaxies and clusters, and the first so-called ``standard siren'' for
cosmography at high redshift well before the advent of LISA and third
generation GW detectors (see Smith et al. 2018b and references
therein).

By 2022 LIGO/Virgo will have reached design sensitivity, and be
detecting $\sim10-100$ binary neutron star (NS-NS) mergers per year
out to a typical distance of $D_{\rm L}\simeq200\,{\rm Mpc}$ (Abbott
et al. 2016; Chen et al. 2017).  In the following few years,
LIGO/Virgo's sensitivity will improve by a further factor $\sim2$
beyond design sensitivity, with typical sources detected at $D_{\rm
  L}\simeq400\,{\rm Mpc}$.  At these distances a source with an
electromagnetic counterpart similar to AT2017gfo/GW170817 will have an
apparent magnitude of ${\rm AB}\simeq24.5$ in $g/r/i$-band filters
within $t_{\rm rest}<3\,{\rm days}$ following GW alert (Arcavi 2018).
In summary, LSST data of depth comparable similar to a single WFD
survey visit will be sensitive enough to find large numbers of
kilonova counterparts to NS-NS mergers that are not lensed well in to
the 2020s.  LSST follow-up observations of NS-NS mergers that are not
lensed are discussed in white papers from DESC and TVSSC.

We forecast a rate of detection of lensed NS-NS mergers of $\sim0.01$
per Earth year at the start of LSST survey operations, rising to
$\sim0.1$ per Earth year in the first few years of the LSST survey, as
LIGO/Virgo's sensitivity improves beyond their current design
sensitivity (Robertson et al. 2019).  The typical lensed NS-NS merger
will be located at $z_{\rm true}\simeq2$, and will be inferred by
LIGO/Virgo to be a merger between two black holes (BHs) of mass
$M_{\rm LIGO}=1.4(1+z_{\rm true})/(1+z_{\rm LIGO})\simeq4M_\odot$ at
$z_{\rm LIGO}\simeq0.1$ ($D_{\rm L,LIGO}\simeq600\,{\rm Mpc}$; we
refer to Smith et al. 2018b for details of how lensing affects
interpretation of LIGO/Virgo data).  In summary, the lens
magnification suffered by GW sources cancels the inverse square law
because the strain amplitude measured by LIGO/Virgo, $A$, goes as:
$A\propto\mu^{0.5}{D_{\rm L}}^{-1}$, where $\mu$ is the gravitational
magnification.  Therefore the brightness of the counterpart to a
lensed GW source is set by the luminosity distance initially inferred
by LIGO/Virgo when assuming $\mu=1$.  Therefore an AT2017gfo-like
optical counterpart to a typical lensed NS-NS merger will have the
same apparent magnitude as a similar source located at $D_{\rm
  L,LIGO}=600\,{\rm Mpc}$: ${\rm AB}\simeq-13.5+5\log(600\,{\rm
  Mpc})-5=25.4$.  In short, lensed GW science is impossible with data
of WFD single-visit depth, because the transient nature of the targets
requires them to be detected in a single epoch of observations, and
the most sensitive filter will reach ${\rm AB}\lesssim24.5$ in a
single WFD visit.

We therefore advocate creating a LSST Target of Opportunity (ToO)
programme that is flexible enough to allow a small number of ToO
follow-up observations of candidate lensed NS-NS that are deeper than
WFD depth.  We also advocate that the WFD survey commences with a
reference survey of the entire observable extragalactic sky, with
sufficient depth in $g$- and $i$-bands to enable it to be used as the
reference imaging for discovery of lensed NS-NS/kilonovae.

\section{Technical Description}\label{sec:tech-intro}

\subsection{High-level description}\label{sec:gw-discovery}

Our high-level requirements assume that candidate lensed NS-NS mergers
are located at $z_{\rm true}=2$ and are interpreted by LIGO/Virgo
(assuming $\mu=1$) as being at $z_{\rm LIGO}\simeq0.1$, i.e. $D_{\rm
  L,LIGO}\simeq600\,{\rm Mpc}$.  When estimating the apparent
magnitude of kilonova counterparts, we assume that the absolute
magnitude of AT2017gfo and other kilonovae associated with short GRBs
(Gompertz et al. 2018) are typical of kilonovae counterparts of high-z
NS-NS mergers.  We also conservatively assume that these candidates
will be localized to a $100\,{\rm degree}^2$ uncertainty by
LIGO/Virgo.  This is the median forecast localization uncertainty of
GW detections relevant throughout the 2020s, as we wait for new GW
detectors (e.g. LIGO-India) to come online (Abbott et al. 2016).
Based on these assumptions, the discovery of kilonova counterparts to
multiply-imaged NS-NS mergers requires the following:
\begin{itemize}
\item \textbf{ToO imaging triggered by LIGO/Virgo events, with each
  ToO spanning a contiguous area of 100 degree$^2$ (12 pointings).}
  LSST is the only machine capable of surveying $\sim100\,{\rm
    degree}^2$ down to ${\rm AB}\simeq25.4$ within a timescale of a
  few nights following a GW alert;
\item \textbf{Three epochs of observations per ToO through two filters
  (preferably $g$- and $i$-band, but $g$- and $r$-bands also
  acceptable)} to constrain the colour and temporal evolution of
  candidate kilonovae, and thus exclude non-kilonova transients from
  further investigation (Cowperthwaite et al. 2018).
\end{itemize}

\subsection{Footprint -- pointings, regions and/or constraints}

For strongly lensed NS-NS/kilonovae, the sky location and detailed
design of the pointing strategy will depend on the details of an
individual NS-NS sky localization from LIGO/Virgo.  The sky
localizations will be randomly distributed on the sky.

\subsection{Image quality}

We have no strong constraint on image quality for the first epoch of
observations of lensed NS-NS ToOs, although sub-arcsecond is
preferred.  In the second and third epoch (preferably within two days
of, and acceptable up to one week after GW alert), we prefer image
quality no worse than the first epoch, however the timing constraint
is more important than this preference.

\subsection{Individual image depth and/or sky brightness}

There will be no major constraints on image depth or sky brightness
for the first epoch, because the main driver is timeliness — the first
epoch must be as fast as possible after the GW alert.  The second and
third epochs must be within one week of the GW alert, and therefore
could allow some flexibility in scheduling, image quality (see above),
and sky brightness, in that we prefer conditions no worse than
achieved in the first epoch.  This would maximise the probability of
re-detecting a transient source in data of the same integration time
as the first epoch.

\subsection{Co-added image depth and/or total number of visits}

We request three epochs per candidate lensed NS-NS merger.  Each epoch
consists of $100\,{\rm degree}^2$ imaging in two filters filters, with
typically $6\times30$ second exposures per filter per epoch.  This
depth is required to reach ${\rm AB}\simeq25.4$.  More precise target
depths and numbers of exposures will be possible in response to
specific candidates.  The numbers given here are representative.

\subsection{Number of visits within a night}

Three epochs are required, with each epoch being completed within a
single night.  Each epoch comprises 12 pointings, observed through 2
filters.  If it is not possible to complete an epoch within the night,
then on any given night we prefer to observe a subset of the pointings
through both filters in order to obtain contemporaneous colour
information.  In this situation, the remaining pointings should be
completed on the following night.

\subsection{Distribution of visits over time}

\subsubsection{Discovery of lensed NS-NS/kilonovae}

The first epoch must be as fast as possible after the GW alert,
preferably on the first available Chilean night.  The second epoch is
preferred on the first night following the GW alert, several hours
after the first epoch.  The third epoch is preferred on the second
night following the GW alert.  This optimal strategy will give the
necessary colour and temporal information to distinguish kilonovae
from other sources such as supernovae.

It is also important to be realistic, given weather and other
scheduling constraints.  The time dilation suffered by the source at
$z\simeq2$ therefore works to our advantage, as we can tolerate the
first epoch being on the second night following GW alert, and the
latter two epochs being spaced further apart, up to a maximum of
$\sim7$ nights following GW alert, before the putative kilonova fades
from view.

\subsubsection{A Deep and Early All-sky Reference Imaging Survey}\label{sec:reference}

In addition to a LSST ToO programme, discovery of lensed
NS-NS/kilonovae requires a reference image of the entire extragalactic
sky available to LSST as early in the survey as possible.  It is
desirable that this reference image would include $g$- and $i$-band
imaging down to ${\rm AB}\simeq25.5$ in the early months of the
survey.  This depth is $\sim1\,{\rm mag}$ fainter than the nominal
single-visit depth, and will therefore require $\sim6$ visits per
filter.  This ``deep and early all-sky reference survey'' strategy
will deliver reference imaging sensitive enough to support the
detection of a kilonova counterpart to a lensed NS-NS right from the
beginning of the LSST survey. We select the $g$- and $i$-bands to
provide a broad wavelength range for colour information on candidate
kilonovae without straying in to the less sensitive $u/z/y$-bands.

A less ambitious ``all-sky reference imaging'' strategy would be a key
requirement for general strong lensing science for a variety of
cases. This strategy would ensure that all the available extragalactic
sky observed in the $g$- and $i$-bands to single-visit depth at the
beginning of the survey operations.  This would be very powerful
because single visit $g$- and $i$-band depth is sufficient for
discovery of galaxy/group/cluster-scale strong lenses.  Knowing where
all the bright lenses early on in the WFD survey will allow us to
identify lensed kilonovae/NS-NS quickly.  This early census of lenses
spanning galaxy- to cluster-scales will also be invaluable to
constrain the run of optical depth to strong-lensing with lens mass,
and thus to refine the observing and analysis strategies when
searching for lensed GWs.

\subsection{Filter choice}

It is essential that we observe through two filters, and we prefer
$g$- and $i$-band filters to maximize the wavelength range over which
we will be able to obtain colour information, without sacrificing
sensitivity by going to $u/z/y$-band filters.  This colour information
is required to exclude non-kilonova transient sources, as discussed
for example by Cowperthwaite et al. (2018).

\subsection{Exposure constraints}

No constraints.

\subsection{Other constraints}

We request a flexible ToO mode be implemented in the general LSST
observing strategy.  Optimally, we request ToO observations of
$\sim100\,{\rm degree}^2$ down to ${\rm AB}\simeq25.4$, deeper than
the single visit depth. Given that these will be areas that would be
covered by the WFD strategy anyway, a mechanism for triggering
(modified) observations from the standard WFD survey tiles that would
encompass the full LIGO/Virgo localization and meet the depth and
cadence requirements outlined in this White Paper should be explored.
This would allow the ToO observations to also be used as part of the
WFD survey, thereby minimizing the impact of interrupting the WFD
observing schedule, and reducing the effective cost of the proposed
ToOs.

\subsection{Estimated time requirement}\label{sec:time}

Based on the assumptions listed in Section 3.1.2, we detail the amount
of observing time required per candidate lensed NS-NS below:
\begin{itemize}
    \item Simultaneous slew to first field and change filter to $i$ if necessary: 120 seconds
    \item Visits to all 12 pointings in the $i$-band: (30sec $+$ 3sec slew/settle $+$ 2 sec shutter open/close) $\times$ 12 $\times$ 6 $=$ 2520 seconds
    \item Change filter to $g$-band: 120 seconds
    \item Repeat observing in $g$-band: 2520 seconds
    \item Return LSST system to its previous state (simultaneous slew and filter change): 120 seconds
\end{itemize}
This totals 5400 seconds, i.e. just 1.5 hours, per epoch.  For three
epochs per trigger, this equates to 4.5 hours per trigger.  At an
estimated rate of $\sim0.1$ trigger per Earth year, we therefore
expect one or a few such triggers over the 10 year duration of the
LSST survey, say 15 hours of ToO observations, which is just $0.06\%$
of the total time (3650 nights x 0.83 uptime x 8 hours/night = 24236
hours) available.


\begin{table}[ht]
    \centering
    \begin{tabular}{|l|l|}
        \toprule
        Properties & Importance \\
        \midrule
        Image quality  &  2 \\
        Sky brightness &  3 \\
        Individual image depth & 2 \\
        Co-added image depth & 2 \\
        Number of exposures in a visit   & 1 \\
        Number of visits (in a night)  & 1 \\ 
        Total number of visits & 1 \\
        Time between visits (in a night) & 2\\
        Time between visits (between nights)  & 1\\
        Long-term gaps between visits & 3\\
        \bottomrule
    \end{tabular}
    \caption{{\bf Constraint Rankings:} Summary of the relative
      importance of various survey strategy constraints. Please rank
      the importance of each of these considerations, from 1=very
      important, 2=somewhat important, 3=not important. If a given
      constraint depends on other parameters in the table, but these
      other parameters are not important in themselves, please only
      mark the final constraint as important. For example, individual
      image depth depends on image quality, sky brightness, and number
      of exposures in a visit; if your science depends on the
      individual image depth but not directly on the other parameters,
      individual image depth would be `1' and the other parameters
      could be marked as `3', giving us the most flexibility when
      determining the composition of a visit, for example.}
        \label{tab:obs_constraints}
\end{table}


\subsection{Technical trades}

There are no obvious trades for lensed NS-NS/kilonova discovery because they are based on ToO observations.


\section{Performance Evaluation}

Our science goals will be achieved if the full sequence of
observations across three epochs is completed within the required
timescales.

Each ToO trigger will be scored on a scale from 0 to 36.  The score is
incremented by 1 when a pointing has been observed through both
filters for the required $6\times30$ seconds within the required time
constraints ($<2$ days from GW alert for epoch 1, $<4$ days from GW
alert for epoch 2, and $<7$ days from GW alert for epoch 3).
Successful completion of one pointing in all three epochs therefore
scores 3.  With 12 pointings required to observe the full $100\,{\rm
  degree}^2$ sky localisation, the maximum score is 36.  We define a
ToO as successful if it scores $\ge27$ on this system.

We will design the layout of the pointings such that they prioritise
observations of the central regions (higher probability) of the
LIGO/Virgo sky localisation over the lower probability regions.  In
this way, the impact of partial completion of the ToO programme on
statistical inferences relating to detection of a kilonova counterpart
can be estimated in a meaningful way.


\section{Special Data Processing}

Data processing will be the same as the standard LSST pipeline except
that the 6 exposures per filter per pointing per epoch will need to be
stacked before subtracting the relevant reference images.  A custom
filter may be required to run on the resulting alert stream, that
detects candidate lensed kilonovae based on their colour, time
evolution, and proximity to individual early-type galaxies or
groups/clusters of galaxies that are acting as the foreground lens.


\section{References}

Abbott et al., 2016, Living Reviews in Relativity, 19, 1\\
Arcavi et al., 2018, ApJ, 855, L23\\
Chen et al., 2017, arXiv:1709.08079\\
Cowperthwaite et al., 2018, arXiv:1811.03098\\
Gompertz et al., 2018, ApJ, 860, 62\\
Robertson A., Smith G.~P., et al., 2019, in prep., and available here soon:\\ ~~~~\url{http://www.sr.bham.ac.uk/~gps/preprints.html}\\
Smith G.~P., Jauzac M., et al., 2018a, MNRAS, 475, 3823\\
Smith G.~P., Bianconi M., et al.\ 2018b, arXiv:1805.07370\\

\end{document}